\documentstyle[11pt, aaspp4, psfig]{article}

\def\gta{\ifmmode {\mathbin{\lower 3pt\hbox   
    {$\,\rlap{\raise 5pt\hbox{$\char'076$}}\mathchar"7218\,$}}}
    \else {${\mathbin{\lower 3pt\hbox
    {$\rlap{\raise 5pt\hbox{$\char'076$}}\mathchar"7218\,$}}}
    $}\fi}
\def\lta{\ifmmode {\,\mathbin{\lower 3pt\hbox   
    {$\,\rlap{\raise 5pt\hbox{$\char'074$}}\mathchar"7218\,$}}}
    \else {${\mathbin{\lower 3pt\hbox
    {$\rlap{\raise 5pt\hbox{$\char'074$}}\mathchar"7218\,$}}}
    $}\fi}

\begin{document}

\title{Reionization Constraints on the Contribution of Primordial Compact 
Objects to Dark Matter}

\author{M. Coleman Miller}
\affil{Department of Astronomy, University of Maryland\\
       College Park, MD  20742-2421\\
       miller@astro.umd.edu}

\begin{abstract}

Many lines of evidence suggest that nonbaryonic
dark matter constitutes $\sim$30\% of the critical closure density, but
the composition of this dark matter is unknown.  One class of candidates
for the dark matter is compact objects formed in the early universe, with
typical masses $M\sim 0.1-1\,M_\odot$ to correspond to the mass scale 
of objects found with microlensing observing projects.  Specific candidates
of this type include black holes formed at the epoch of the QCD phase
transition, quark stars, and boson stars.  Here we
show that accretion onto these objects produces substantial
ionization in the early universe, with an optical depth to Thomson
scattering out to $z\sim 1100$ of 
$\tau\approx 2-4 [f_{\rm CO}\epsilon_{-1}(M/M_\odot)]^{1/2}
(H_0/65)^{-1}$, where $\epsilon_{-1}$ is the accretion efficiency
$\epsilon\equiv L/{\dot M}c^2$ divided by 0.1 and $f_{\rm CO}$ is 
the fraction of matter in the
compact objects.  The current upper limit to the scattering optical
depth, based on the anisotropy of the microwave background, is $\approx 0.4$.
Therefore, if accretion onto these objects is relatively efficient, 
they cannot be the main component of nonbaryonic dark matter. 
\end{abstract}


\section{Introduction}

Observations of the rotation curves of galaxies and clusters, in addition 
to joint fits of Type~Ia supernova data and
the power spectrum of the cosmic microwave background, suggest that
the density of matter in the current universe is $\sim$30\% of the
closure density, i.e., $\Omega_m\sim 0.3$.  However, the success 
of big bang nucleosynthesis in explaining the primordial abundances of 
light elements, especially the primordial abundance ratio of D/H, requires 
that the contribution of baryons is only $\Omega_b h^2=0.019\pm 0.0024$
(95\% confidence; Tytler et al.\ 2000), where 
$h\equiv H_0/100$~km~s$^{-1}$~Mpc$^{-1}$ and $H_0$ is the present day
Hubble constant.  The majority of the matter must be something else.

One class of possibilities involves hypothesized
exotic particles, from light particles such as axions (Peccei \& Quinn 1977) 
to heavier particles such
as the neutralino (e.g., Jungman, Kamionkowski, \& Greist 1996) 
or even ultramassive
particles such as ``WIMPZILLAs" (Kolb, Chung, \& Riotto 1998;
Hui \& Stewart 1999).
Another class, which we focus on in this paper, involves dark matter
that occurs primarily in $\sim 0.1-1\,M_\odot$ clumps.  This class,
which has received recent attention because this is the mass scale
of objects discovered by microlensing projects such as MACHO, EROS,
and OGLE, has several specific candidates.  For example, black holes
may have formed during the QCD phase transition from quark matter to
nucleonic matter (Jedamzik 1997, 1998; Niemeyer \& Jedamzik 1999), 
during which the horizon mass was plausibly
in the $0.1-1\,M_\odot$ range.  Other suggestions involve quark
stars (Banerjee et al.\ 2000), boson stars 
(Colpi, Shapiro, \& Wasserman 1986; Mielke
\& Schunck 2000), and stars formed of mirror matter (Mohapatra \&
Teplitz 1999).
Here we consider those members of this class that involve primordial
compact objects, specifically those objects which (1)~existed before
the $z\sim 1100$ epoch of decoupling, and (2)~have a mass to radius
ratio of $GM/Rc^2\gta 0.1$.  These include black holes, quark stars,
and boson stars, but not mirror matter stars, as
they are envisioned to form at comparatively late times 
and to be comparable to ordinary stars in their compactness
(Mohapatra \& Teplitz 1999).

Primordial compact objects
will accrete from the ambient medium and will therefore generate substantial
luminosity.  This luminosity can ionize the surrounding medium.  Unlike
the energy spectra from ordinary stars, which drop off rapidly above
the ground state ionization energy of hydrogen, the energy spectra from
accreting compact objects are known observationally to be very hard,
with substantial components above 1~keV and often extending above
100~keV.  An important consequence of this is that whereas the Stromgren
sphere of ionization around, say, an O or B star is extremely sharply
defined, with an exponentially decreasing ionization fraction outside
the critical radius, the ionization fraction produced by an accreting compact
object dies off relatively slowly with radius, as $r^{-3/2}$ (Silk 1971;
Carr 1981).  Therefore, accretion onto a primordial object can produce 
ionization over a large volume in the early universe.  If the resulting 
optical depth to Thomson scattering is too large, it will conflict with the
upper limit to this optical depth derived from the observed anisotropy
of the microwave background (Griffiths, Barbosa, \& Liddle 1999).
Conversely, the upper limit on the optical depth can be used to
constrain the properties of primordial compact objects, if these are
proposed as the dominant component of dark matter.

Here we calculate the ionization produced by compact objects accreting
in the early universe.  We find that the ionization produced by secondary
electrons, an effect not included in previous analyses of reionization
by accretion, increases substantially the ionization fraction and hence
the optical depth to Thomson scattering.  In \S~2 we
show that the Thomson optical depth out to the
$z\approx 1100$ redshift of decoupling is $\tau\approx 2-4 
[f_{\rm CO}\epsilon_{-1}(M/M_\odot)]^{1/2}
(H_0/65)^{-1}$, where $\epsilon_{-1}$ is the accretion efficiency
$L/{\dot M}c^2$ divided by 0.1 and $f_{\rm CO}$ is the fraction of matter in
primordial compact objects.  We compare this result to the
current observational upper limit of $\tau<0.4$, and show that
either low accretion efficiency or low mass is required if dark matter is
mostly composed of primordial compact objects.  In \S~3 we consider
low-efficiency accretion such as flows dominated by advection or wind
outflow.  We show that the constraints from ionization are especially tight
on objects without horizons.
In \S~4 we place this result in the context of previous constraints on, for
example, primordial black holes as the main component of dark matter.
We also discuss future improvements to our result.  In particular,
we show that the expected accuracy of optical depth measurements with MAP
and Planck could decrease the upper bound on $f_{\rm CO}\epsilon_{-1}
(M/M_\odot)$ by a further factor of $\sim$100.

\section{Calculation of Optical Depth}

If the number
density of baryons is $n(z)\approx z^3n_0$ (where in this entire
calculation we assume $z\gg 1$) and the ionization fraction is
$x(z)$, then the optical depth to Thomson scattering between redshifts
$z$ and $z+dz$ is
\begin{equation}
d\tau(z)=n(z)x(z)\sigma_T ds(z)
\end{equation}
where $\sigma_T=6.65\times 10^{-25}$ is the Thomson scattering
cross section and
\begin{equation}
ds(z)={1\over H_0}{cdz\over{(1+z)E(z)}}
\end{equation}
is the distance traveled by a photon in this redshift interval.  Here 
$E(z)=\left[\Omega_m(1+z)^3+\Omega_R(1+z)^2+\Omega_\Lambda\right]^{1/2}$
and $\Omega_m$, $\Omega_R$, and $\Omega_\Lambda$ are the current
contributions to the mass energy of the universe from, respectively,
matter, curvature, and the cosmological constant.  At $z\gg 1$ the
first term dominates, so that $E(z)\approx \Omega_m^{1/2}z^{3/2}$ and
$ds(z)\approx cH_0^{-1}\Omega_m^{-1/2}z^{-5/2}dz$.  The differential
optical depth is then
\begin{equation}
d\tau(z)\approx n_0z^3x(z)\sigma_T ds(z)=n_0x(z)\sigma_T
{c\over{H_0 \Omega_m^{1/2}}}z^{1/2}\,dz
\end{equation}
(see also Haiman \& Knox 1999).
This needs to be integrated out to the $z\sim 1100$ redshift of decoupling
to determine the optical depth to scattering in the early universe.
The main unknown in this expression is the ionization fraction $x(z)$.
In the remainder of this section, therefore, we compute the ionization
produced by radiation from accreting compact objects.  In \S~2.1 we
compute the luminosity and spectrum of this radiation.  We assume
Bondi-Hoyle accretion and a spectrum corresponding to that observed
from many neutron stars and black hole candidates.  In \S~2.2 we use the
ionization balance equation to calculate the ionization produced by a
single source.  We include the effects of ionization by secondary electrons,
which is a significant effect not included in the analysis of 
pregalactic black hole accretion by Carr (1981).
In \S~2.3 we show that the ionizing flux from sources spread throughout
the universe increases significantly the ionization fraction.  Finally, in
\S~2.4 we calculate the optical depth to Thomson scattering out
to the $z\sim 1100$ redshift of decoupling, including the effects of
Compton cooling by the microwave background.

\subsection{Luminosity and Spectrum of Radiation}

Let us now consider accreting objects of mass $M$.  
Suppose that these masses are
moving with the Hubble flow, so that the main parameter governing the
accretion rate is the speed of sound in the gas at infinity, 
$a_\infty=\sqrt{\Gamma_1kT/\mu m_p}$, where $\Gamma_1$ is the polytropic
index, $m_p$ is the mass of the proton, 
and $\mu$ is the mean molecular weight.  For pure hydrogen ($\mu=1/2$), 
the mass accretion rate from a perfect gas with $\Gamma_1=5/3$ is then
\begin{equation}
{\dot M}=1.2\times 10^{10}\left(M\over M_\odot\right)^2
\left(\rho_\infty\over{10^{-24}{\rm g\ cm}^{-3}}\right) 
T_4^{-3/2}\,{\rm g\ s}^{-1}\; ,
\end{equation}
where $T_4\equiv T_\infty/10^4$~K and $T_\infty$ is the temperature of
the gas at infinity.  For a primordial composition of 75\% hydrogen
and 25\% helium by mass, this accretion rate is more than doubled
because helium has half the velocity for a given temperature that
hydrogen does, and hence accretes at eight times the rate for a given
mass density.  We therefore take the coefficient to be $3\times 10^{10}$.
If accretion produces a luminosity with an efficiency $0.1\epsilon_{-1}$,
so that $L=10^{20}\epsilon_{-1}{\dot M}$~erg~s$^{-1}$, then
\begin{equation}
L=3\times 10^{30}\left(M\over M_\odot\right)^2
\left(\rho_\infty\over{10^{-24}{\rm g\ cm}^{-3}}\right)
T_4^{-3/2}\,{\rm erg\ s}^{-1}\; .
\end{equation}
The best estimate of the baryon density in the current universe
from big bang nucleosynthesis constraints (Tytler et al.\ 2000) is
\begin{equation}
\rho_{B0}=3.6\pm 0.4\times 10^{-31}\,{\rm g~cm}^{-3}\; .
\end{equation}
At a redshift $z$ this density is therefore $(1+z)^3\rho_{B0}\approx
z^3\rho_{B0}$.
Hence, if the compact object accretes matter with the
average baryonic density in the universe, the luminosity at redshift $z$ is
\begin{equation}
L\approx 10^{24}z^3\left(M\over M_\odot\right)^2
T_4^{-3/2}\,{\rm erg\ s}^{-1}\; .
\end{equation}
Pressure balance of a hot HII region with the cooler exterior universe
may decrease the density of accreting matter and therefore decrease
this luminosity (see below).  In accreting black hole sources from
stellar mass to AGN, and also in some accreting neutron stars, the
spectrum often has a power-law tail with equal power in equal
logarithmic intervals of the photon energy, up to some $E_{\rm max}$:
$dL(E)/dE\propto E^{-1}\exp(-E/E_{\rm max})$.  The results of our
calculation are fairly insensitive to the assumed spectrum.  Normalizing this
spectrum so that the total luminosity above $E_0=13.6$~eV is $L$, 
the differential
photon flux at energy $E$ a distance $R$ from the compact object is
\begin{equation}
F(E)=e^{-\tau(E)}{dL/dE\over{4\pi R^2 E}}=
e^{-\tau(E)}{L\over{4\pi\ln(E_{\rm max}/E_0)E^2 R^2}}e^{-E/E_{\rm max}}\; .
\end{equation}
Here $\tau(E)$ is the optical depth at a distance $R$ from the source
to photons of energy $E$.  Note that Carr (1981) chose a spectrum of
a bremsstrahlung form ($dL/dE\propto e^{-E/E_{\rm max}}$), and
hence had a different energy dependence and normalization for the photon 
number flux.

\subsection{Secondary Ionization and Ionization Balance}

A given photon can effectively produce many ionizations, because the
ionized electrons can collisionally ionize other atoms (see, e.g., 
Silk \& Werner 1969; Silk 1971).  The collisional cross section
exceeds $10^{-17}$~cm$^2$ for electron energies between
$\sim$15~eV and 1~keV (see Dalgarno, Yan, \& Liu 1999 for a recent
discussion of electron energy deposition).  Dalgarno et al.\ (1999)
calculate that the mean energy per ion pair decreases with increasing
initial electron energy, reaching a limit of 36.1~eV per pair at
energies $>$200~eV.  Using their Figure~6, we adopt an approximate
value of $E/3E_0$ hydrogen atoms ionized by a photon of initial energy
$E$; this is a rough average over the energy range of interest, and we
assume for simplicity that it is constant over that range.

The effective ionization rate produced by the photons generated by 
accretion is therefore (adapting the formula of Carr 1981)
\begin{equation}
\zeta_H\approx \int_{E_0}^\infty \sigma_1\left(E\over E_0\right)^{-3}
\left(E\over{3E_0}\right)F(E)\,dE\; ,
\end{equation}
where $\sigma_1\approx 2\times 10^{-17}$~cm$^2$ is
the abundance-weighted ionization cross section at $E_0$.
The integrand in this formula is a constant factor
\begin{equation}
{E_{\rm max}\over{3E_0\ln(E_{\rm max}/E_0)}}
\end{equation}
times the integrand in the corresponding formula in Carr (1981).  The
difference arises because we assume a different form for the spectrum
and account for the ionization produced by secondary electrons.
The remainder of the analysis 
of the ionization region created by a single source follows the treatment of
Carr (1981), with this factor included.  This is a large factor, of
order 25 for $E_{\rm max}=10^{-8}$~erg, and it therefore makes a
crucial difference to the overall ionization.  If $E_{\rm max}\gg E_0$,
the ionization rate is approximately (Silk et al.\ 1972)
\begin{equation}
\zeta_H\approx \left(E_{\rm max}\over{3E_0\ln(E_{\rm max}/E_0)}\right)
{\sigma_1L\over{12\pi E_{\rm max}R^2}}
\left[1-\exp(-\tau_0)\over{\tau_0}\right]\; ,
\end{equation}
where 
\begin{equation}
\tau_0=\int_0^R n_H(1-x)\sigma_1\,dR
\end{equation}
and the first factor in parentheses indicates the correction factor
to the expression of Carr (1981).  Here $x$ is the ionized fraction
at radius $R$.

The ionization balance equation is
\begin{equation}
\alpha n_H^2 x^2=\zeta_H n_H(1-x)\; ,
\end{equation}
where around $T\approx 10^4$~K, the recombination coefficient 
not including single-photon transitions to the ground state
(which would release ionizing photons) is
$\alpha\approx 2.6\times 10^{-13}T_4^{-0.75}$~cm$^3$~s$^{-1}$ 
(Hummer 1994).  Far from the accreting compact
object, where $x\ll 1$ and $\tau_0\gg 1$, the ionization fraction is
\begin{equation}
x=\left(E_{\rm max}\over{3E_0\ln(E_{\rm max}/E_0)}\right)^{1/2}
{1\over{\sqrt{8}}}\left(R\over{R_s}\right)^{-3/2}\; ,
\end{equation}
where
\begin{equation}
R_s=\left[2L\over{3\pi\alpha n_H^2 E_{\rm max}}\right]^{1/3}\; .
\end{equation}
Again, the initial factor in the equation for $x$ is the correction
factor, which therefore increases the ionization fraction far from
the compact object by a factor of $\sim$5; note that the only remaining
dependence on $E_{\rm max}$ is $\left[\ln(E_{\rm max}/E_0)\right]^{-1/2}$,
so this result is very insensitive to the high-energy cutoff of the
spectrum.

\subsection{Contribution of Multiple Sources}

The total ionization rate $\zeta_H$ must be summed over the contributions
of all sources.  At large distances from a source, $\tau_0\propto R\gg 1$,
so that $\zeta_H\sim R^{-3}$.  For multiple sources separated by an average 
distance $R_{\rm sep}$, the ionizing rate is larger than the single-source
ionizing rate at a distance $R_{\rm sep}$ by a factor
\begin{equation}
{\sum\zeta_H\over{\zeta_H(r=R_{\rm sep})}}=\int_{R_{\rm sep}}^{R_{\rm max}}
\left(r\over{R_{\rm sep}}\right)^{-3}4\pi r^2 n_{\rm CO}\,dr\; .
\end{equation}
Here $n_{\rm CO}=10^{-62}z^3(M/M_\odot)^{-1}\Omega_{\rm CO}$~cm$^{-3}$
is the number density of compact objects at redshift $z$, where
$\Omega_{\rm CO}$ is the fraction of the closure density in compact
objects.  Also, $R_{\rm max}\approx {\rm min}\left[10^{31}z^{-3}x^{-1},
c/H(z)\right]$~cm is the mean
free path to Thomson scattering.
The separation distance is approximately given by
$\left({4\over 3}\pi R_{\rm sep}^3\right)^{-1}=n_{\rm CO}$, so
$R_{\rm sep}^3\approx {3\over{4\pi}}n^{-1}_{\rm CO}$.
Therefore, 
\begin{equation}
{\sum\zeta_H\over{\zeta_H(r=R_{\rm sep})}}\approx 3\ln(R_{\rm max}/
R_{\rm sep})\; .
\end{equation}
The ratio of radii is typically $10^6-10^8$, so the enhancement due to
the contributions of multiple sources is approximately a factor of 50.

When multiple sources are included, the ionization fraction (for $x\ll 1$)
is increased by a factor that is approximately the square root of the
factor by which the ionization rate is enhanced.  At a distance $R$
the rate is enhanced by a factor
\begin{equation}
\zeta_H\rightarrow \zeta_H\left(1+50(R/R_{\rm sep})^3\right)\; ,
\end{equation}
and hence the ionization fraction including multiple sources is
\begin{equation}
x\approx \left(E_{\rm max}\over{3E_0\ln(E_{\rm max}/E_0)}\right)^{1/2}
{1\over{\sqrt{8}}}\left(R\over{R_s}\right)^{-3/2}
\left(1+50(R/R_{\rm sep})^3\right)^{1/2}\; .
\end{equation}
Integrating this from $R_s$ to $R_{\rm sep}$, the volume-averaged
ionization is 
\begin{equation}
{\bar x}\approx 3\left(E_{\rm max}\over{3E_0\ln(E_{\rm max}/E_0)}\right)^{1/2}
\left(R_s\over{R_{\rm sep}}\right)^{3/2}\; .
\end{equation}
Here $R_{\rm sep}=3\times 10^{20}z^{-1}(M/M_\odot)^{1/3}
\Omega_{\rm CO}^{-1/3}$~cm.

\subsection{Optical Depth Including Compton Cooling and Pressure Balance}

An evaluation of this expression for the ionization fraction requires
knowledge of the luminosity of individual sources and the average temperature
of the matter in the universe.  As pointed out by, e.g., Carr (1981), the
dominant cooling process at high redshifts is inverse Compton cooling
off of the microwave background.  If the temperature $T$ of the matter
is much larger than the temperature $T_r$ of the radiation background,
then the cooling rate per volume at redshift $z$ is
\begin{equation}
\Gamma_r\approx 2\times 10^{-38}x(z)z^7 T_4\ {\rm erg\ cm}^{-3}
{\rm s}^{-1}\; .
\end{equation}
The average heating rate is just the luminosity per source times the
number density of sources:
\begin{equation}
\Gamma_h=Ln_{\rm CO}=10^{-38}\Omega_{\rm CO}z^6\ {\rm erg\ cm}^{-3}
{\rm s}^{-1}\; .
\end{equation}
At $z\sim 1000$, where the optical depth to Thomson scattering exceeds
unity and as we will see $x\sim 0.1$, the cooling rate
dominates the heating rate and hence the matter temperature is locked
to the radiation temperature during this epoch (see also Carr 1981).
This increases the recombination rate over most of the volume of
interest, and therefore decreases the optical depth to scattering.
Inside the HII region, by contrast, heating dominates cooling and the
temperature remains close to $10^4$~K for $z>10$; in fact, Carr (1981)
finds that the temperature is $T_4=(z/10^3)^{0.3}$.  The
temperature difference means that pressure balance requires that the
density inside the HII region be less than the average density by a
factor $\sim {\bar T}/T$; note, however, that this
configuration requires the support against gravity of a denser by a
less dense medium, which therefore is in principle Rayleigh-Taylor unstable.
Hence, mixing could occur which would decrease the temperature and increase
the density of the HII region.  If mixing does not occur, the density
of the matter accreting onto the compact object is decreased by a factor 
$\sim 0.27\,(z/10^3)^{0.7}$.
The ionization fraction and hence the optical depth would therefore be
reduced by the square root of this factor, or about $0.5\,(z/10^3)^{0.35}$.  
The uncertainty of whether there is an interchange instability and mixing
thus produces an uncertainty of a factor $\sim$2 in the optical depth to
scattering.

With these contributions, the average ionization is
\begin{equation}
{\bar x}(z)\approx 5-10\times 10^{-4}\epsilon_{-1}^{1/2}
\left(M\over M_\odot\right)^{1/2}
\Omega_{\rm CO}^{1/2}\left[6\over{\ln(E_{\rm max}/E_0}\right]^{1/2}z^{0.7}\; .
\end{equation}

The differential optical depth to scattering is $d\tau(z)=
n_0 x(z)\sigma_T {c\over{H_0\Omega_m^{1/2}}}z^{1/2}\,dz$, so using
$n_0=2.2\pm 0.3\times 10^{-7}$~cm$^{-3}$ and evaluating the constants,
this is
\begin{equation}
d\tau(z)\approx 1-2\times 10^{-6}\left[6\over{\ln(E_{\rm max}/E_0)}\right]^{1/2}
\epsilon_{-1}^{1/2}(M/M_\odot)^{1/2}
(H_0/65\,{\rm km}\,{\rm s}^{-1}\,{\rm Mpc}^{-1})^{-1}
f_{\rm CO}^{1/2}z^{1.2}\,dz\; .
\end{equation}
Here $f_{\rm CO}\equiv\Omega_{\rm CO}/\Omega_m$ is the fraction of 
matter in compact objects.
Integrating from a small redshift to the redshift $z\approx 1100$ at
decoupling gives finally 
\begin{equation}
\tau\approx 2-4 \left[{6\epsilon_{-1}\over{\ln(E_{\rm max}/E_0)}}
\left(M\over M_\odot\right)f_{\rm CO}
\right]^{1/2} \left(H_0\over{65}\right)^{-1}\; .
\end{equation}

The effect of this optical depth on the observed CMB power spectrum is
not identical to the effect of the same optical depth if it came from
sudden and complete reionization at some lower redshift $z\sim 10-40$.
The reason is that the mechanism described here produces most of the
optical depth at comparatively high redshifts, $z\gta 800$, and hence
for optical depths in excess of unity the scatterings occur close to
recombination where some of the primordial anisotropy is maintained.
In contrast, scattering at low redshift exponentially suppresses the
primordial anisotropy.  However, if the optical depth is less than
unity this effect is less pronounced in the reionization mechanism
discussed in this paper, because scatterings occur over a wide range
of redshift and hence tend to smooth out small-scale anisotropies in
the same way as would happen due to scattering at much lower redshifts.
These qualitative effects are confirmed by simulations with CMBFAST
(Seljak \& Zaldarriaga 1996 and subsequent papers), which show that
for $\tau\lta 1$ the constraints on the optical depth from the observed
CMB power spectrum are roughly the same for this mechanism as for
late reionization.

The observational upper limit to $\tau$ from small-scale CMB anisotropy
is $\tau\lta 0.4$ if $\Omega_m=0.3$ and the primordial power spectrum
has an index $n=1$ (Griffiths et al.\ 1999).  To be consistent with
this limit, primordial compact objects must therefore be either
low-efficiency accretors, low-mass objects, or a minor component of
dark matter.  Given that the measured mass spectrum of MACHOs in our galaxy
has a peak in the $\sim 0.5\,M_\odot-1\,M_\odot$ range (Alcock et al.\ 2000;
note, however, that most of the mass in the halo need not be in
MACHOs [Gates et al.\ 1998, Alcock et al.\ 2000] and a higher-mass 
component is not ruled out [Lasserre et al.\ 2000]), 
explanation of these objects with a population that composes most of the
dark matter in the universe requires low-efficiency accretion, which we
consider in the next section.  These constraints are particularly strict
for higher-mass black holes.  The joint
limits on $M$ and $\epsilon$ are shown in Figure~1, for three different
upper limits to the Thomson optical depth: $\tau$=0.4, 0.14, or 0.05, 
which are the optical depths obtained if the the universe was fully
and suddenly reionized at a redshift of $z_{\rm reion}$=40, 20, or 10,
respectively.

\begin{figure}
\psfig{file=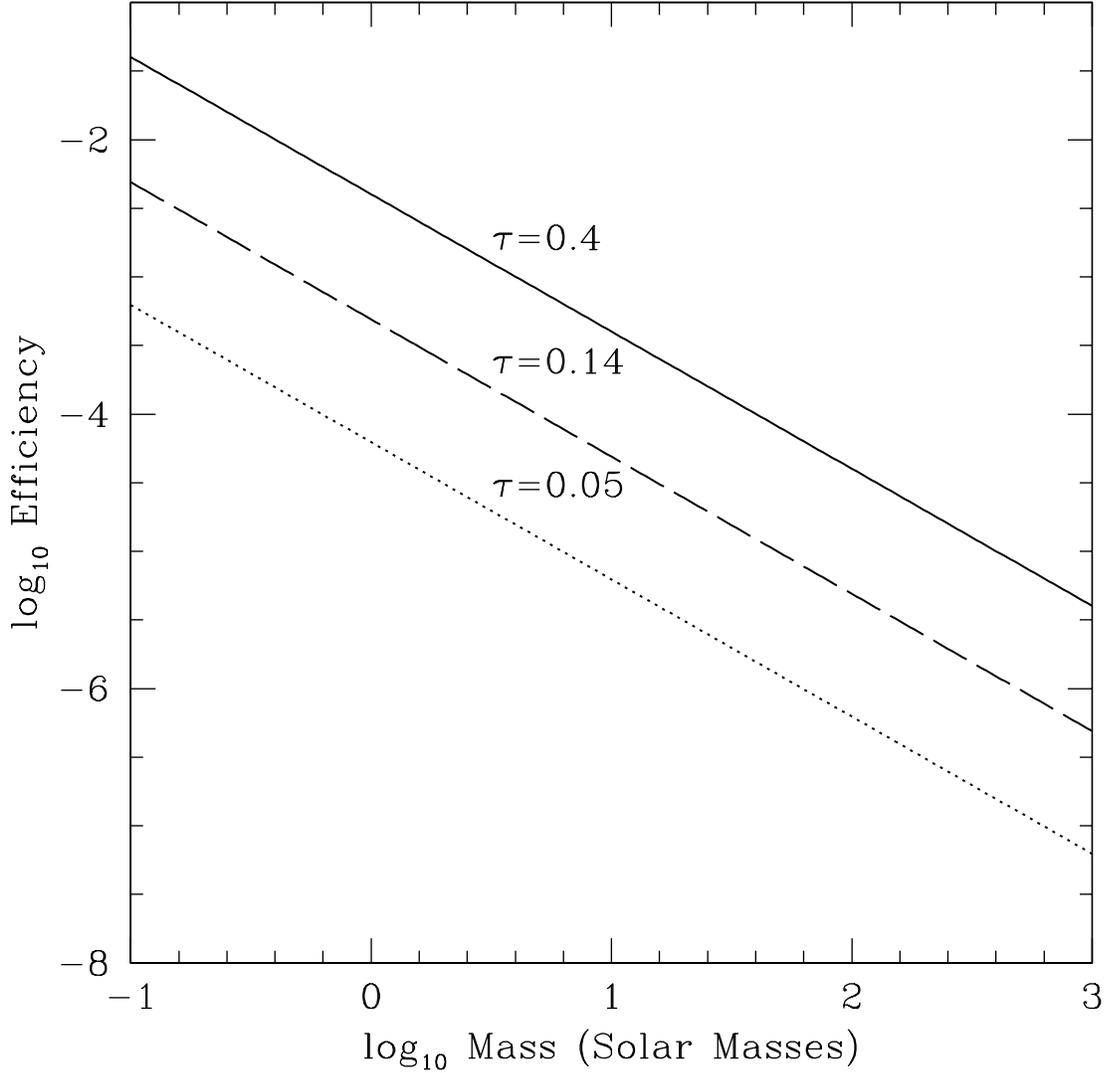,height=6.0truein,width=6.0truein}
\caption{Limits on the accretion efficiency $\epsilon=L/{\dot M}c^2$ of
primordial compact objects as a function of their typical mass, if they are 
to make up the bulk of dark matter ($f_{\rm CO}=1$).
These limits assume no Rayleigh-Taylor mixing at the boundary of
the HII region, and are therefore conservative.  The three
curves are the upper limits to the efficiency as a function of mass
for the current upper limit to the Thomson optical depth of
$\tau=0.4$ (solid curve) and future possible upper limits of
$\tau=0.14$ (dashed curve) and
$\tau=0.05$ (dotted curve).}
\end{figure}

\section{Efficiency of Accretion}

In the last few years there has been much discussion of the
possibility that, for low accretion rates, an advection-dominated
accretion flow (ADAF) is set up in which the radiative efficiency onto
black holes is low because the matter flows  almost radially into the
hole, taking almost all of its energy with it (e.g., Narayan \& Yi
1994; Abramowicz et al.\ 1995). The radiative efficiency according to
these solutions could be extremely low, and hence might allow a large
matter density in black holes.  If the compact object does not have a
horizon, then an ADAF will not reduce the radiative efficiency, so
this is not a way out.

Another possibility is that most of the accreting matter does not reach
the surface at all, perhaps because it is driven out in a wind.  
This is the basis of the advection-dominated
inflow-outflow solution (ADIOS) proposed by Blandford \& Begelman
(1999). Convective flow, in which the net inward flow rate at small
radii is small, has also been found analytically  by Quataert \&
Narayan (1999) and in numerical simulations by Stone, Pringle, \&
Begelman (1999).  If in Bondi-Hoyle accretion the result of the
accretion  is inflow of a small fraction of matter combined with
outflow of most of the matter, then the accretion efficiency onto even
objects without horizons could be small.  However, there is a crucial
unsolved problem with these flows, which is whether they will remain at
low efficiency indefinitely if there is a steady inflow of matter from
infinity, as in Bondi-Hoyle accretion.  If instead the accretion
proceeds as in a dwarf nova, in which there is a long-term buildup of
matter followed by a short-term, high-luminosity episode during which the
accumulated matter is dumped onto the central object, then current
accretion theory suggests the accretion will generate radiation
efficiently regardless of the nature of the compact object.  In such a
case, neither black holes nor any other type of primordial compact
object are viable candidates for most of the dark matter in
the universe.

\section{Discussion and Conclusions}

Consideration of compact objects as components of dark matter
has often been restricted
to black holes, but many of the arguments apply more generally.  Black
holes with masses in excess of $\sim 10^3\,M_\odot$ are ruled out as a
significant component of galactic halos because their dynamical
interactions with globular clusters would destroy the clusters (for a
recent calculation see Arras \& Wasserman 1999).  The lack of an increase
in the number of low equivalent width quasars with increasing redshift
(expected to be caused by gravitational lensing)
rules out a contribution $\Omega\gta 0.1$ from any objects with
masses between $\sim 10^{-2}\,M_\odot$ and $20\,M_\odot$ that are more
compact than their Einstein radii (Dalcanton et al.\ 1994).  The lack
of observed lensing of cosmological gamma-ray bursts also allows weak
limits to be placed on the contribution of black holes of various sizes:
$\Omega<0.15$ at the 90\% level for $M=10^{6.5}\,M_\odot$, $\Omega<0.9$
at the $1\,\sigma$ level for $M=10^{-12.5}-10^{-9}\,M_\odot$, and
$\Omega<0.1$ ($z_{\rm GRB}\sim 1$) or $\Omega<0.2$ ($z_{\rm GRB}\sim 2$)
at the 95\% level for $M=10^{-16}-10^{-13}\,M_\odot$ (Marani et al.\ 1999).

Here we show that ionization from compact object accretion in the 
early post-decoupling universe is more significant than had been thought
previously, because of the effects of secondary ionization by electrons.
The result is that, barring
inefficient accretion ($\epsilon<0.05$ for $M=0.1\,M_\odot$, 
$\epsilon<0.005$ for $M=1\,M_\odot$), primordial compact objects in
this mass range cannot compose a significant fraction of the mass of
the universe, because they would ionize the universe enough to conflict
with the measured small-scale anisotropies of the cosmic microwave
background.  If further
analysis and numerical simulation of flows onto black holes demonstrates
that the long-term time averaged accretion efficiency is $\gta 0.1$,
as might happen if matter tends to pile up as in a dwarf nova and then
accrete quickly with efficient radiation, then all masses greater than
$\sim 0.1\,M_\odot$ are excluded from making a significant contribution.

Future CMB missions such as MAP and Planck could strengthen
these constraints considerably.  The optical depth resolution of MAP
is expected to be 0.022, and of Planck is expected to be 0.004 
(Zaldarriaga, Spergel, \& Seljak 1997; Bouchet, Prunet, \& Sethi 1999; 
Eisenstein, Hu, \& Tegmark 1999).
Since the existence of a Ly$\alpha$ emitter at $z=5.64$ (see Haiman
\& Spaans 1999) shows that
reionization must have occurred before then, this means that both
satellites, and especially Planck, will be able to detect the effects
of ionization regardless of the actual redshift of reionization.  If
$z_{\rm reion}\sim 10$ then the
redshift of reionization could even be determined directly with SIRTF
or NGST via, e.g., analysis of the damping wing
of the Gunn-Peterson trough (Miralda-Escud\'e 1998) or detection of
transmitted flux between Lyman resonances (Haiman \& Loeb 1999).  The
upper limit on the product $\epsilon_{-1}(M/M_\odot)f_{\rm CO}$ scales
like $\tau_{\rm scatt}^2$ (or, for $z\gg 1$, like $z_{\rm reion}^3$), 
so if $z_{\rm reion}\sim 10$ this upper limit is
decreased by almost a factor of 100.  In this case, barring extremely
inefficient accretion, dark matter must be composed of less 
compact objects or of WIMPs.

\acknowledgements
We thank Sylvain Veilleux, Andy Young, Jim Stone, Eve Ostriker, and
Scott Dodelson for comments.  This work was supported in part by NASA 
ATP grant number NRA-98-03-ATP-028.

\end{document}